\documentclass{emulateapj}
\usepackage{amsmath}
\usepackage{color}

\shorttitle{The metallicity of NGC~4038}
\shortauthors{Lardo et al.}

\begin{document}

\title{Red Supergiants as cosmic abundance probes: the first direct metallicity determination of 
NGC~4038 in the Antennae}

\author{C.~Lardo\altaffilmark{1}, B.~Davies\altaffilmark{1}, R-P.~Kudritzki\altaffilmark{2,3}, J.~Z.~Gazak\altaffilmark{2}, C.~J.~Evans\altaffilmark{4}, L.~R.~Patrick\altaffilmark{5}, M.~Bergemann\altaffilmark{6}, B.~Plez\altaffilmark{7}}
\altaffiltext{1}{Astrophysics Research Institute, Liverpool John Moores University, IC2, Liverpool Science Park, 146 Brownlow Hill, Liverpool, L3 5RF, UK}
\altaffiltext{2}{Institute for Astronomy, University of Hawaii, 2680 Woodlawn Drive, Honolulu, HI, 96822, USA}
\altaffiltext{3}{University Observatory Munich, Scheinerstr. 1, D-81679, Munich, Germany}
\altaffiltext{4}{UK Astronomy Technology Centre, Royal Observatory Edinburgh, Blackford Hill, Edinburgh EH9 3HJ, UK}
\altaffiltext{5}{Institute for Astronomy, University of Edinburgh, Royal Observatory Edinburgh, Blackford Hill, Edinburgh EH9 3HJ, UK}
\altaffiltext{6}{Max-Planck Institute for Astronomy, 69117, Heidelberg, Germany}
\altaffiltext{7}{Laboratoire Univers et Particules de Montpellier, Universit\'{e} de Montpellier, CNRS, F-34095 Montpellier, France}

\begin{abstract}
We present a direct determination of the stellar metallicity in the close pair galaxy NGC~4038 (D = 20 Mpc) 
based on the quantitative analysis of moderate resolution KMOS/VLT  spectra of three super star clusters (SSCs).
The method adopted in our analysis has been developed and optimised
 to measure accurate metallicities from atomic lines in the $J$-band of single red supergiant (RSG) or RSG-dominated star clusters.
Hence, our metallicity measurements are not affected by the biases and poorly understood systematics inherent to {\em strong line} \ion{H}{2} methods  which are routinely applied to massive data sets of galaxies.
We find [Z]= +0.07 $\pm$ 0.03 and  compare our measurements to \ion{H}{2} strong line calibrations.
Our abundances and literature data suggest the presence of
a flat metallicity gradient, which can be explained as redistribution of metal-rich gas following the strong interaction.

\end{abstract}


\section{Introduction}

The metallicity of a galaxy is moderated by the cycling of chemically processed material by stars and any 
gas exchange between the galaxy and the environment.
The central metallicity is correlated with a galaxy mass, a relation 
which holds information about galaxy formation and 
evolution \citep{lequeux79,tremonti04,maiolino08,kudritzki12}.
Additionally, the variation of the metallicity of the galaxy 
with the distance from the centre
keeps track of the complex dynamics of galaxy evolution,
as several fundamental physical processes affect metallicity gradients
\citep[e.g.][]{searle71,zaritsky94,garnett97,prantzos00,chiappini01,fu09,pilkington12,mott13,kudritzki15}.
Obtaining reliable metallicities in galaxies is crucial to measure and interpret the behaviour of metallicity 
with radial distance and the mass-metallicity relation. 
Unfortunately, robust metallicity measurements in 
galaxies are notoriously problematic to obtain. Metallicity of starburst or star-forming galaxies is routinely measured 
from \ion{H}{2} region
emission lines and two main methods are employed: the T$_{\rm{e}}$-based method and the {\em strong line} method.
The T$_{\rm{e}}$-based method  uses the flux ratio of auroral to strong lines of the same ion to measure the electron temperature of the gas
 \citep{rubin94,lee04,stasinska05,andrews13}.  However, 
temperature sensitive lines are often too weak to be detected in faint distant galaxies
and their measurement is challenging even for galaxies in the local universe specifically in the 
metal-rich regime \citep{stasinska05,bresolin05,ercolano10,zurita12,gazak15}. 
When the electron temperature cannot be determined, one has to resort to abundance indicators based on more readily observable lines. 
Such {\em strong line} methods are based on the ratio of the fluxes of the strongest forbidden lines of typically O and H \citep{pagel79,skillman89,mcgaugh94} which are more easily detected than the weak auroral lines across a wide range of metallicity.
Unfortunately, well-known but unexplained systematic uncertainties, that can amount up to  $\sim$0.7 dex,
plague the determination of the metallicity of extragalactic \ion{H}{2} regions from nebular spectroscopy \citep{kewley08,bresolin09}.
As a consequence, the chemical abundances derived from strong-line methods display large systematic differences when applied to the same observational data; and only relative metallicity comparisons
appear to be reliable if the same calibration is used.

Very promising tracers of the present-day abundances in star-forming galaxies are evolved massive stars.
Indeed, a number of nearby galaxies have been studied using blue supergiants (BSGs) and 
results indicate excellent agreement with abundances obtained from the T$_{\rm{e}}$-based method in 
\ion{H}{2} regions (see \citealp{kudritzki12, kudritzki13,kudritzki14,hosek14} and references therein).

A growing body of evidence indicates that accurate metallicities over large 
distance scales as for BSGs can be derived also from red supergiant stars (RSGs).
RSGs are young ($\leq$ 20 Myr) and extremely bright stars, i.e. 10$^5$-10$^6$ L$_{\sun}$ \citep{humphreys79}. 
Their flux peaks at $\simeq$ 1$\mu$ m, therefore they are among the most luminous objects in a galaxy in the near-IR and 
are ideal candidates for directly measuring  extragalactic abundances.
RSGs are also very cool stars, with temperatures ranging from 3000 to 5000 K, hence their spectra 
show numerous absorption features \citep{allard00}.
Previous techniques to measure metallicity from RSG spectra concentrated on the $H$-band, where 
high resolution (R[$\lambda / \delta \lambda$]$\simeq$ 20000) observations are needed to isolate 
diagnostic atomic lines from molecular absorption \citep[e.g.][]{davies09}. Even with the largest available telescopes, 
the need for high resolution translates prohibitively large  
exposure times for individual objects at a distance $\geq$1 Mpc.
In contrast, the required observing time can be significantly reduced if one 
focuses on a narrow region in the $J$-band, where 
the dominant spectral features are {\em isolated atomic lines} of Fe and the $\alpha$ elements (Ti, Si, and Mg).
Indeed, in this spectral window accurate abundances can be measured even at moderate resolution (R[$\lambda / \delta \lambda$]$\simeq$ 3000).
 
The $J$-band method was initially introduced by \citet{davies10} for individual RSGs in the Milky Way and has
been extended and largely tested by \citet{gazakPerOB1} in the association Perseus OB-1.
\citet{davies15} checked the validity of this method at lower metallicity in the Magellanic Clouds using the VLT/XShooter  and 
\citet{patrick15} accurately tailored the reduction/analysis method for KMOS observations of RSGs in NGC~6822. 
\citet{gazak15} obtained metallicities from RSGs across the disk of NGC~300, a spiral galaxy beyond the Local Group, 
finding a striking agreement with the metallicities 
recovered from BSG stars and \ion{H}{2}-region auroral line measurements. 

Interestingly, this technique can be 
also applied to unresolved star clusters rather than individual stars \citep{gazakM83,gazakSSC}.
In merging and starburst galaxies intense star formation activity triggers the formation of super star clusters (SSCs), 
agglomerates of millions of young ($\leq$ 50 Myr) stars.  
SSCs have masses $\geq$ 10$^5$M$_{\sun}$ and are extremely compact (with radii $\leq$ 5 pc; \citealp{portgeis10}).
Once a SSC reaches an age of $\simeq$ 7 Myr, the most massive stars which have not yet exploded as supernovae will be in the RSG phase. 
For a cluster with an initial mass of 10$^5$ M$_{\sun}$, there may be more than a hundred 
 RSGs present which dominate the cluster's light output in the near-IR, contributing 90\% - 95\% of the of the near-IR flux. As their spectra are all very similar in the effective temperature (T$_{\rm{eff}}$) range around 4000~K, the combined spectrum can be analysed in the same way  as a single RSG spectrum, as shown by \citet{gazakSSC}.  Therefore, for SSCs older than 7 Myr the $J$-band technique 
 can be used to measure metallicity at far greater distances than is possible for single supergiants. 
 
Following this line of investigation, \citet{gazakSSC} analysed two young SSCs in 
M~83 (at 4.5 Mpc,  \citealp{thim03}) and NGC~6946 (at 5.9 Mpc, \citealp{kara00}), 
finding metallicities $\simeq$1.5-2.0 and $\simeq$0.5 $\times$  solar,
respectively. This paper further extends the observational baseline and presents quantitative
$J$-band spectroscopy of three SSCs in the close pair galaxy NGC~4838.

NGC~4038 is the main component of 
the Antennae system (NGC~4038/39), the closest ($\simeq$20 Mpc)\footnote{\citet{riess11} estimated a distance modulus to the Antennae galaxies
of m-M=31.66 $\pm$ 0.08 ($\simeq$ 22.3 Mpc) from optical and infrared observations of Cepheid variables with the 
the Wide Field Camera 3 (WFC3) on the Hubble Space Telescope (HST).}
and youngest example of an ongoing major merger, involving two gas-rich disk galaxies that began to collide $\simeq$ 200-400 Myr ago \citep{barnes88,mihos93}. 
Galaxy mergers, and their resulting starbursts, are one of the basic building blocks of structure formation in the universe (e.g., \citealp{baron87}) and 
represent an ideal laboratories for close up studies of the physical processes 
that were important  at the peak of cosmic star formation \citep{deravel09,bundy09}. 
As such, the Antennae has, over the years, been the favourite target for several multi wavelength studies of the effect 
of tidal interaction (e.g., \citealp{whitmore99,fabbiano04,hibbard05,gilbert07,brandl09,whitmore10,klaas10,
whitmore14}) and numerous N-body and hydrodynamical simulations (e.g., \citealp{toomre72,barnes88,mihos93,teyssier10,karl10}).

The bodies of the galaxies are sites of extensive star formation ($\sim$20 M$_{\sun}$  yr$^{-1}$; \citealp{ zhang01})
 producing an IR luminosity of log L$_{\rm{IR}}$ = 10.76, which is  is an order of magnitude lower than ultraluminous infrared galaxies, but 
 still a factor of $\sim$5 higher than noninteracting galaxy pairs (see, e.g., \citealp{kennicutt87}).
Most of the star formation in this colliding galaxy pair occurs in the form of SSCs (\citealp{whitmore95,mirabel98,whitmore99,wilson00}), 
with masses up to a few times 10$^6$M$_{\sun}$ which are distributed throughout the galaxy \citep{zhang01}.

Observational studies demonstrate that in interacting galaxies, the strong metallicity gradient observed in isolated 
spirals, can be disrupted by gas flows of metal-poor gas from the outer regions towards the centre of the galaxy \citep[e.g.][]{kewley10}.

Here we analyse KMOS/VLT  spectra of three SSCs 
to measure the central metallicity and the metallicity gradient  across the disk of NGC~4038. 
This paper is organised in the following way. In Section~\ref{OBSERVATIONS}, we summarise the observations and data reduction.  We outline our analysis procedure in Section~\ref{ANALYSIS}. 
Our results are presented in Section~\ref{RESULTS}. We summarise our main conclusions in Section~\ref{SUMMARY}.

\begin{figure*}\label{fig1}
\centering
\includegraphics[width=16cm]{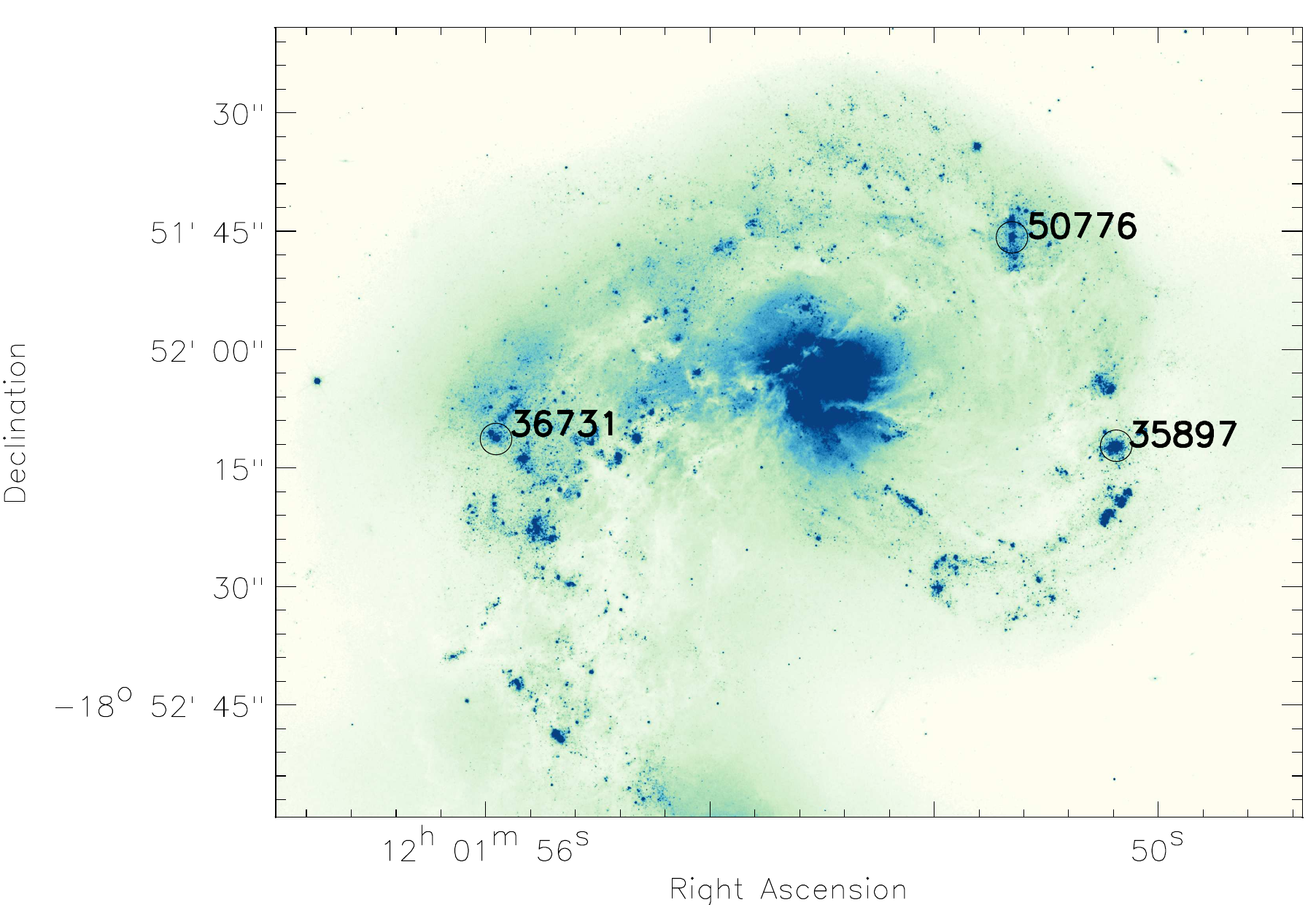}
\caption{The targeted SSCs are overlayed over a HST-ACS F814W image of NGC~4038 retrieved from the Hubble Legacy Archive ({\tt http://hla.stsci.edu/hlaview.html}).}
\end{figure*}

\section{Target selection, observations and data reduction}\label{OBSERVATIONS}

\begin{deluxetable*}{cccccccc}
\tabletypesize{\scriptsize}
\tablecaption{SSCs targeted. Data are from \citet{whitmore10} \label{TAB1}}

\tablewidth{0pt}
\tablehead{
\colhead{ID} & \colhead{R.A.} & \colhead{Decl.} & \colhead{M$_{V}$}  & \colhead{m$_{J}$\tablenotemark{(a)}} &
 \colhead{E($B - V$) } &
\colhead{Log$\tau$/yr} & \colhead{Mass (M$_{\sun})$}}
\startdata
35897 & 12:01:50.4453 & -18:52:14.223 & -11.6   & 15.8 &  0.00  & 7.6	& 4.5 $\times$10$^5$ \\
36731 & 12:01:55.9896 &	-18:52:12.985 & -12.5  & 17.3 &  0.04   & 7.6	& 1.1 $\times$10$^6$  \\
50776 & 12:01:51.3963 &	-18:51:47.562 & -14.4  & 16.4 & 0.04   & 6.8	& 1.1 $\times$10$^6$  \\
\enddata
\tablenotetext{(a)}{J magnitudes are from aperture photometry on archival Wide Field Camera 3 images. 
The typical uncertainty on the J magnitude is $\sim$ 0.1 mags. }

\end{deluxetable*}

\begin{figure}
\epsscale{1.0}
\centering
\includegraphics[width=8cm]{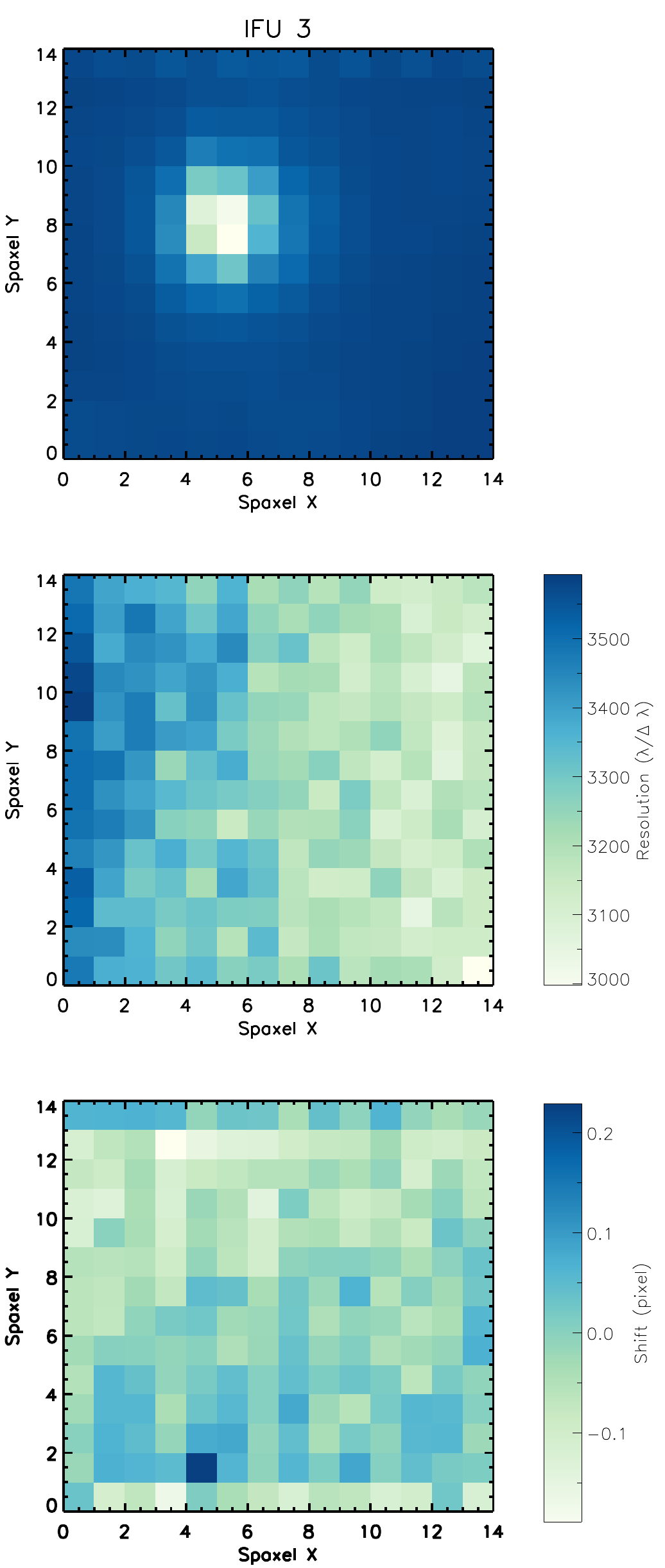}
\caption{Variation of spectral resolution and wavelength calibration across the spatial
pixels of a typical KMOS IFU measured from night sky lines.
The top panel presents the reconstructed image obtained in a 300 sec exposure.
The middle and bottom panels show the variation of spectral resolution  and 
the wavelength shifts (in pixel)  across the spaxels, respectively.}
\label{fig:kmog}
\end{figure}

Our sample consists of three RSG dominated SSCs, whose coordinates, luminosities, colours, ages 
are listed in Table~\ref{TAB1} together with other useful information.
Target SSCs with the appropriate luminosity and colours were selected from optical and near-IR photometry by 
\citet{whitmore10} from the Advanced Camera for Surveys (ACS) and the Near Infrared Camera and Multi-Object Spectrometer (NICMOS) 
mounted on the HST. Their spatial location across the galaxy is shown in Figure~\ref{fig1}.
The observations were carried out with KMOS/VLT \citep{sharples13} in April 2014 in visitor mode (PI: Kudritzki: \dataset{093.B-0023}),
with a total exposure time of 6000s split in 20 subsequent exposures.

KMOS is a spectrograph equipped with 24 deployable integral-field units (IFUs)
that can be allocated within a  7.2\arcmin~diameter field-of-view (FoV). 
Each IFU covers a projected area on the sky of about 2.8\arcsec $\times$ 2.8\arcsec, which is sampled by an array of 
14$\times$14 spatial pixels (spaxels) each with an angular size of 0.2\arcsec.
The 24 IFUs are managed by three identical spectrographs, each
one handling eight IFUs. The observations were performed in nod-to-sky mode with the 
YJ grating, covering the 1.00-1.35
$\mu$m 
spectral range with a nominal resolution of R$\simeq$3600 at the band centre.
 Observations were carried out using the standard AB AB-like object-sky sequence (i.e. one sky frame for each object frame) in which we offset by 5\arcsec~to sky, and each observation was dithered by up to 0.2\arcsec.  During the observations, the average $J$-band seeing 
was approximately 1.0\arcsec.  In addition to science observations, a standard set of calibration frames were obtained.
Since we require high precision absorption line spectroscopy, we observed a telluric standard
with the arms in the science configuration (i.e. using the observational template that 
allows users to observe a standard star in each IFU allocated to a science target).
During the observations 7 SSCs were observed, 
but only three of those have sufficient SNR ($\geq$ 100, see \citealp{gazakPerOB1}) to be used in our analysis.
To reduce the data, we used the standard recipes provided by the by the Software Package for Astronomical Reduction with KMOS (SPARK; \citealp{spark}). KMOS IFU data cubes were flat  fielded, wavelength calibrated, and telluric corrected using the
standard KMOS/esorex routines.

As can be seen in Figure~\ref{fig:kmog}, there are significant variations in both spectral resolution and wavelength calibration across 
the FoV of each IFU as measured from sky emission lines (see below for more details). Left uncorrected, these can introduce sky and telluric cancellation errors into the final spectrum, which can be the source of substantial problems for precision absorption line spectroscopy such 
as that presented in this paper.

We correct for these effects with a process we call, {\it kmogenization} \citep[see also][]{gazak15}. 
We first take the rectified science and sky images {\it prior} to sky subtraction. At each spaxel in the IFU we fit Gaussian profiles to the sky lines and we use them to  determine both the spectral resolution and the higher order wavelength calibration as a function of spatial position. We then smooth the spectra at each position on the IFU down to a lower resolution, set to be $R=3200$\footnote{Though there are regions on the IFU where the spectral resolution is below $R=3200$, the pixels within the extraction aperture are always above this value.}. We then extract the spectrum of the science target from an aperture with radius 1.5 spaxels around the flux peak. This narrow aperture minimises errors due to spatial non-uniformity, at the slight expense of discarding flux from the wings of the point-spread function. When the source is extracted, the wavelength axis is updated to include the higher order correction determined from the sky lines {\it without} interpolating the spectrum onto a new wavelength axis, as this can introduce numerical noise. The associated sky spectrum is extracted in the same way, and is subtracted from the science spectrum.

Since the wavelength calibration can change slightly throughout the night due to the rotation of the instrument, each sky-subtracted science spectrum can have a different wavelength axis. To account for this, the normalised spectra are co-added onto a master wavelength axis whereby the flux at wavelength $\lambda_i$ in the master spectrum is a weighted mean of the fluxes in the individual spectra which have wavelengths between $\lambda_{i-1}$ and $\lambda_{i+1}$. Rejection of outliers is also performed at this stage, to eliminate cosmic rays and bad pixels.

Finally, we experimented with altering the output spectral resolution to test the robustness of our method. We found that the best-fitting parameters (see next section) were stable to well within the fitting errors as long as the output resolution was greater than $R=3100$.

\section{Analysis}\label{ANALYSIS}
Metallicity [Z] (normalised to Solar values, [Z] = log (Z/Z$_{\sun}$)\footnote{We measure Fe, Mg, Si, and Ti abundances from individual lines 
and assume this is representative of the metallicity Z.  While the assumption of a solar-scaled composition appears reasonable, we note that the errors deriving from the assumption of a solar-scaled composition rather than an [$\alpha$/Fe]-enhanced one has a little impact on  
the metallicity measurements, i.e. well within the quoted errors on metallicity.}, effective temperature (T$_{\rm{eff}}$), surface gravity (log $g$) and microturbulence ($\xi$) have been  derived as
extensively explained in the previous papers of this series (e.g. \citealp{davies10,gazakSSC,gazakPerOB1,davies15,patrick15,gazak15}).
In particular, the studies of \citet{davies10}, \citet{gazakPerOB1}, and \citet{gazak15} demonstrate the applicability of the technique to objects with roughly solar chemical enrichment.

Atmospheric parameters and metallicity were derived by comparing the observed spectra with a grid of single-star synthetic spectra
degraded to the same spectral resolution as those observed.
Model atmospheres were calculated with the MARCS code \citep{gustafsson08}, where the range of 
parameters are defined in Table~\ref{TAB2}. 
The synthetic spectra were computed using the updated version of the SIU code, as described in \citet{bergemann12}.
Departures from local thermodynamic equilibrium (LTE) for \ion{Fe}{1},  \ion{Mg}{1}, \ion{Si}{1}, and \ion{Ti}{1} lines were also included \citep{bergemann12,bergemann13,bergemann15}.  All other lines including the weak molecular contributions
are calculated in LTE.
The best fit model has been derived through a $\chi ^{2}$-minimisation between the observed spectrum and a
template spectrum at each point in the model grid, taking into account possible 
shifts and variations in instrumental spectral resolution between the observed spectra and the models.
The methodology for  finding the best-fitting model is described in detail by \citet{gazakPerOB1} and \citet{gazak15} which analyse 11 RSGs in Perseus OB-1 ([Z] = --0.04 $\pm$ 0.08) and 27 RGGs in NGC~300 ([Z] = --0.03 $\pm$ 0.05), respectively.
Briefly, we measured best fit parameters by isolating 2D planes in each parameter pair combination which contain the minimum. 
As a result, two parameters are locked to the best fit values in each of the six slices, while allowing the other two parameters to vary. 
We thus constructed a 2D plane of minimum $\chi ^{2}$ values as a function of the two fixed parameters and draw contours of equal $\chi ^{2}$. This procedure is repeated for all the planes, and  the measurements of each parameter are finally averaged to derive best fit parameters. An example of this is shown in Figure~\ref{figPAR}. Figure~\ref{figPAR} shows the degeneracy between [Z] and log $g$,
while demonstrating how soundly constrained are T$_{\rm{eff}}$ and $\xi$.
We refer again to \citet{gazakPerOB1} and \citet{gazak15} for a thorough discussion of the sensitivity of diagnostic lines to the free parameters (e.g. T$_{\rm{eff}}$, log $g$, $\xi$, and [Z]) and  the complete error analysis.
The best-fitting parameters are listed in Table~\ref{TAB3}, along with their associated uncertainties.

\begin{figure*}
\epsscale{1.0}
\centering
\includegraphics[width=14cm]{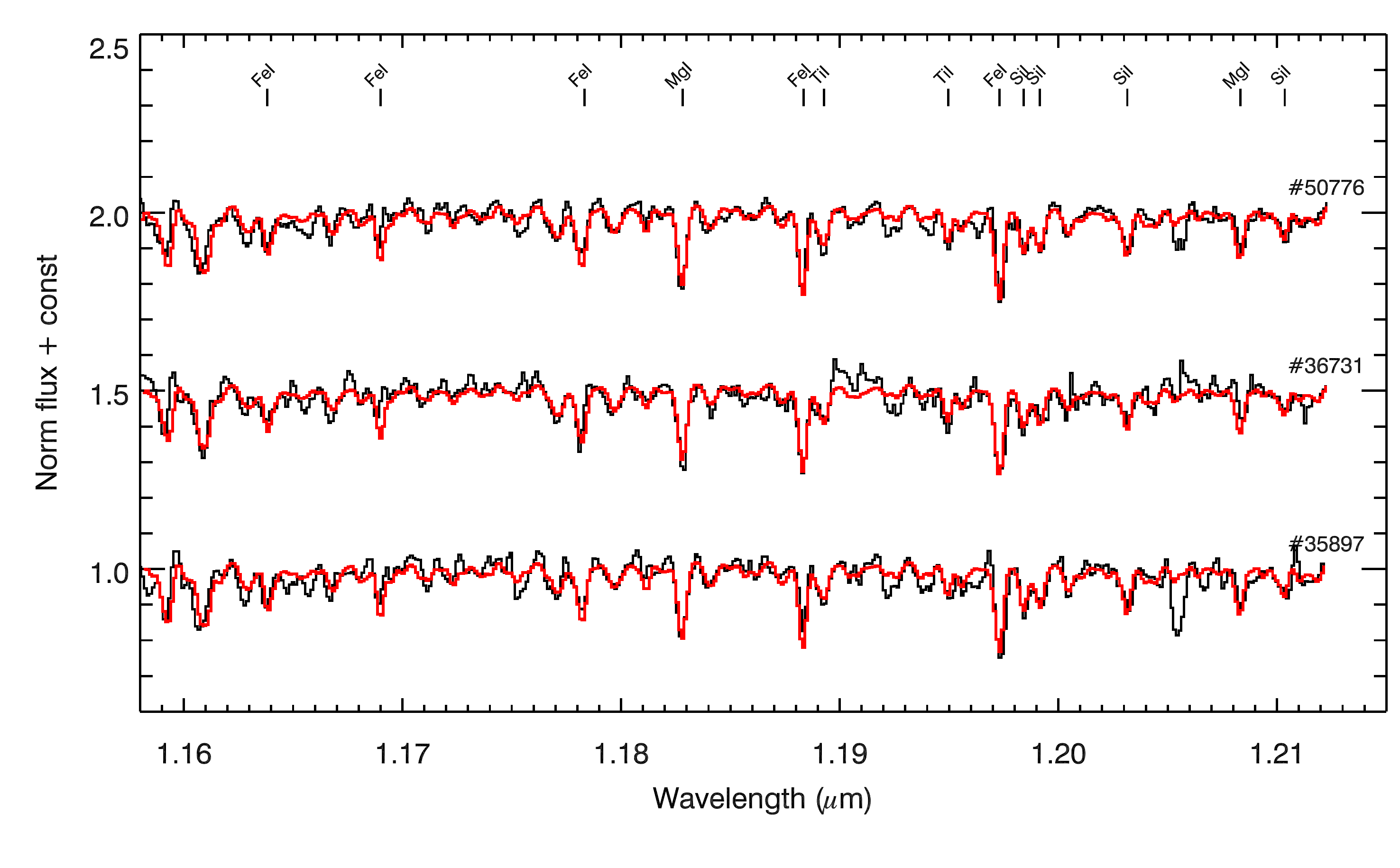}
\caption{Spectra of the targeted SSCs in the $J$-band window, along with the best-fit spectra (red lines). 
}\label{fig:bestfit}
\end{figure*}

\begin{figure}
\centering
\includegraphics[width=\columnwidth]{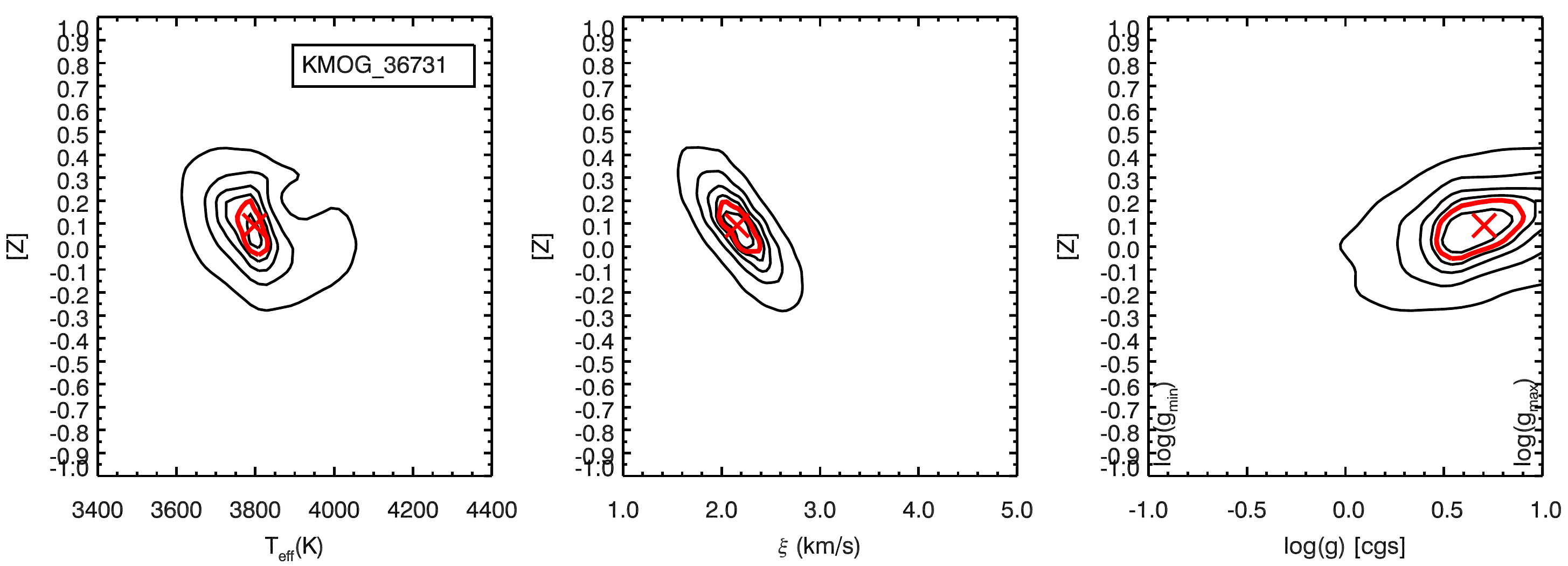}
\caption{Planes through the model grid showing the degeneracy between the parameters for the analysed SSCs .
The black contours show contours of equal $\chi ^{2}$, and are drawn at levels of $\chi ^{2}_{min}$+ (1, 2, 3, 5, 10). 
The red contour is drawn at $\chi ^{2}$ =$\chi ^{2}_{min}$+3 which indicates our 1$\sigma$ uncertainty. 
The $x$-axis limits on the right-hand plot are the minimum and maximum possible values of log $g$ allowed by the
object temperature and luminosity.}\label{figPAR}
\end{figure}

\begin{deluxetable}{crrrr}
\tablecaption{Model grid \label{TAB2} }
\tablewidth{0pt}
\tablehead{
\colhead{Parameter} & \colhead{Units} & \colhead{Grid Min} & \colhead{Grid Max} & \colhead{Grid Step} }
\startdata
T$_{\rm{eff}}$         & K   & 3400 & 4400  & 100\\
$\log$ g	               &dex & --1.0  & + 1.0  & 0.5\\
$[$ Z $]$                           &dex  & --1.00 & +1.00 & 0.25\\
$\xi$                        &km s$^{-1}$ & 1.0 & 5.0 & 1.0\\

\enddata

\end{deluxetable}

\section {Results}\label{RESULTS}
\subsection{Metallicity}
Figure~\ref{fig:bestfit} displays fits of the metal lines with the best fitting model spectra in our diagnostic spectral window.
The match to the stellar lines used to derive metallicity (labelled in the same Figure) is very good, and also the fit to the unresolved 
features of the pseudo-continuum is remarkable. 
The spectral feature at 1.205$\mu$m is due to a poorly removed telluric line.

Table~\ref{TAB3} summarises the derived stellar parameters and metallicities. 
The average metallicity for our sample of 3 SSCs is [Z] = +0.07 $\pm$ 0.03 ($\sigma$=0.05 dex), in
agreement with previous measurements based on different diagnostics.
Solar metallicity has been inferred by \citet{mengel02} from 
the analysis of the metallicity sensitive \ion{Mg}{1} line at 8806.8\AA~in a handful of SSCs.
\citet{bastian09} measure the strong emission lines in \ion{H}{2} regions\footnote{As the lifetimes of RSGs 
are $\leq$50 Myr, their metallicities are expected to be representative of the clouds from which they are formed.} to  
derive metallicities of 16 SSCs, finding a slightly super-solar value (see Figure~\ref{fig3}).
They use the equivalent width ratios of the collisional excited [\ion{O}{2}]$\lambda$3727 and [\ion{O}{3}]$\lambda$4959, 5007 lines relative to the H$\beta$
Balmer series recombination lines to estimate the gas-phase oxygen abundance, i.e. the R23 line ratio.
Moreover, they adopt the analysis method from \citealp{kobu04} (hereafter KK04) and 
a solar abundances of 12+ $\log$(O/H)$_{\sun}$ = 8.89 \citep{edmunds84}, which is 0.2 higher than our adopted solar metallicity,
i.e. 12+ $\log$(O/H)$_{\sun}$ = 8.69 \citep{asplund09}. 
\citet{bastian06} presented a study of several SSCs from VIMOS/VLT spectra. However, they do not have available
the [\ion{O}{2}]$\lambda$3727 emission line strength to measure 
the O/H ratio. Therefore, they use 
 the \citet{vacca92} calibration that relies on the R3 ratio.
 Since this calibration does not include the contribution from another ionised species of oxygen (i.e. [\ion{O}{2}]),
the abundances presented by \citet{bastian06} should be taken with caution, as stated by the authors themselves.
In the following, we do not discuss those abundances further\footnote{Note that two of the SSCs reanalysed in \citet{bastian06} were also observed in \citet{bastian09}, and one on our target (36731) is in common with the \citet{bastian06} sample; i.e.  their complex 3.
For the SSC in common with our sample, \citet{bastian06} estimate a metallicity of Z=0.45 Z$_{\sun}$, with an estimated intrinsic uncertainty of $\pm$0.2 in log(O/H) \citep{edmunds84} and a solar oxygen abundance of 12+ log (O/H)$_{\sun}$=
8.91 \citep{meyer85}, which is compatible with our measurement.}.

\begin{deluxetable}{clrrr}
\tablecaption{Spectral Fits \label{TAB3}}
\tablewidth{0pt}
\tablehead{
\colhead{SSC} & \colhead{T$_{\rm{eff}}$(K)} & \colhead{$\log$ g (dex)} & \colhead{$\xi$ (km/s)} & \colhead{Z (dex)}} 

\startdata
35897 &   3890 $\pm$ 50   &  0.4 $\pm$ 0.2  &  1.7 $\pm$ 0.2  &   0.01 $\pm$ 0.07   \\
36731  &  3750 $\pm$ 50  &  0.6 $\pm$ 0.2   &  2.0 $\pm$ 0.2  &   0.09 $\pm$ 0.08   \\
50776  &  3770 $\pm$ 50  &   0.5 $\pm$ 0.2  &  2.0 $\pm$ 0.2  &    0.11 $\pm$ 0.07   \\
\enddata

\end{deluxetable}

\subsection{Metallicity gradient}
Any sizeable galaxy-galaxy interaction affects the metallicity distribution of the galaxies involved. 
Observational studies of interacting or close pair galaxies have shown that these 
undergo nuclear metal dilution owing to gas inflow, resulting in a significant flattening of their 
gas-phase metallicity gradients \citep{lee04,trancho07,chien07,kewley06,rupke08,ellison08,michel08,kewley10,rich12,sanchez14,rosa14}.

In the last decade, several studies have been published analysing the influence of different levels of interactions in 
the metallicity distribution of  galaxies. 
\citet{kewley10} present the the first systematic analysis of metallicity gradients in close pairs.
They obtain spectra of  star forming regions in eight galaxies in close pair systems
and find that the metallicity gradients are significantly shallower than the gradients in isolated spirals.
\citet{krabbe08} study the kinematics and physical properties of the minor merger AM2306--721. They report
a clear metallicity gradient across the disc of the main galaxy, while the secondary, less massive companion 
shows a relatively homogeneous oxygen abundance. A nearly 
flat radial gradient was measured for both components  of the system AM 2322--821 by \citep{krabbe11}.
\citet{werk11} find that the interacting galaxies in their spectroscopic sample
have flat oxygen abundance gradients out to large projected radii.
A flat metallicity gradient has also been reported for NGC~1512, a barred spiral in a close interaction with a blue compact dwarf companion
\citep{bresolin12} and NGC~92, the largest galaxy in Robert's quartet \citep{torres14}. 
\citet{rosa14} measure oxygen abundance gradients from \ion{H}{2} 
regions located in 11 galaxies in eight systems of close pairs from Gemini/GMOS spectra, 
finding metallicity gradients significantly flatter than those observed in typical isolated spiral galaxies.
Finally, \citet{sanchez14} derive radial gradients of the oxygen abundance in ionised gas in 306 nearby galaxies 
observed by the CALIFA survey. Using a large and homogeneous sample of more than 40 mergers/interacting systems, 
they find statistical evidence of a flattening in the abundance gradients in interacting systems at any interaction stage.

The absence of an abundance gradient in interacting galaxies is explained by invoking efficient mixing  
of low-metallicity gas from the outer parts with the metal-rich gas of the centre of the galaxy.
 Smoothed particle hydrodynamic merger simulations of \citet{barnes06} and \citet{mihos96}
predict that in interacting galaxies the strong tidal interaction
during encounters can aid disk to develop bars.
In such barred disks, low-metallicity gas from the outskirts can efficiently flow towards the central 
regions at higher metallicity funnelled by bar instabilities, flattening the initial radial metallicity gradient 
  \citep{hibbard96,georg00,rampazzo05,rupke05,iono05,emonts06,martin06,cullen07}. 

\begin{figure}
\centering
\includegraphics[width=\columnwidth]{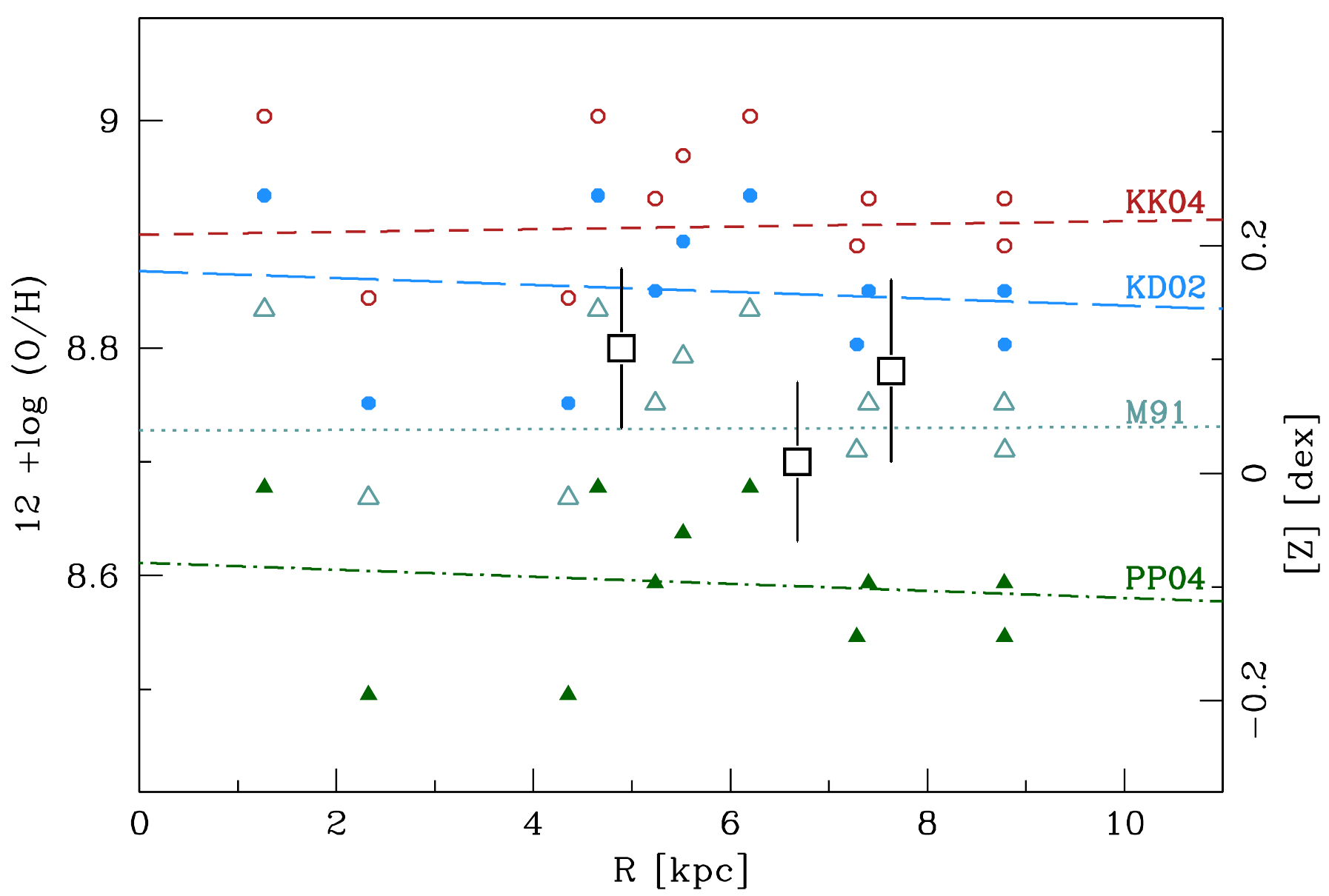}
\caption{Direct metallicities from the SSC analysis (large squares) are shown together with the \ion{H}{2} region metal abundances by 
\citet{bastian09} as a function of galactocentric radius. 
Small symbols are the abundances obtained from different strong-line calibrations to the \citealp{bastian09} data:
\citealp{kobu04} (KK04; red empty circles, adopted by \citealp{bastian09}), \citealp{kewley02} (KD02; blue filled circles),
\citealp{mc91} (M91; teal empty triangles), and \citealp{pettini04} (PP04; green filled triangles).
}\label{fig3}
\end{figure}

We plot our direct metallicity measurements against the radial distance from the centre in Figure~\ref{fig3}.
Galactocentric radii are computed using the direct distance to the galaxy centre, where the galaxy centre is defined using Two Micron All Sky Survey (2MASS) images. Distances take inclination into account ($i$=$56.56^{\circ}$) using the optical diameters from \citet{lauberts89}.  A weighted linear regression to our SSC data yields a central metallicity of +0.17 $\pm$ 0.23 dex and a flat gradient (--0.02 $\pm$ 0.03 dex kpc$^{-1}$), consistent with observations and N-body simulations of interacting systems.

\subsection{Strong-line Abundances} 

Even though the original observations do not include all the different metallicity indicators, we can still compare 
our direct SSC metallicities to strong-line calibrations, as applied to the \citet{bastian09} data\footnote{\citet{bastian09} 
assumed that all \ion{H}{2} regions belong to the upper branch of the R23 calibration.}. 
SSCs metallicities refer to the combined abundances Mg, Si, Ti, and Fe (see Section~\ref{ANALYSIS}, where the individual abundances are scaled 
relative to the solar abundance pattern). In comparing the metallicities from the SSCs with the oxygen abundances from the \ion{H}{2} regions, we assume 
that the oxygen abundances of the SSCs scale with metallicity, and that the solar metallicity value corresponds to 12 + log(O/H)$_{\sun}$ = 8.69 \citep{asplund09}. 

Figure~\ref{fig3} illustrates how four different strong-line calibrations, as applied to the \citet{bastian09} sample,
compare to our metallicities. We consider  the R23 = ([\ion{O}{2}]$\lambda$3727 + [\ion{O}{3}]$\lambda \lambda$4959,5007)/H$\beta$ indicator, with the theoretical predictions by \citealp{mc91} (M91) and KK04, i.e. the calibration used by \citet{bastian09}, the calibration  for the [\ion{N}{2}]$\lambda$6583/[\ion{O}{2}]$\lambda$3727 diagnostic presented in \citealp{kewley02} (KD02), and O3N2 = $\log$ ([\ion{O}{3}]$\lambda$5007/H$\beta$)/([\ion{N}{2}$]\lambda$6584/H$\alpha$), empirically calibrated on \ion{H}{2} regions with T$_{e}$-based metallicities by \citealp{pettini04} (PP04). 
The estimated accuracy of the all these calibrations is $\simeq$0.10 - 0.15 dex.
We use the coefficients of the polynomial from Table~3 in  \citet{kewley08}  to convert metallicities from the KK04 calibration 
into a metallicity that is consistent with other calibrations using a third-order polynomial. 

The abundances that result from the application of these different indicators and calibrations, together with the direct SSC abundances derived in Section~\ref{ANALYSIS}, are shown as a function of the distance from the centre of NGC~4038 in Figure~\ref{fig3}. A linear least-square fit to the data is also shown for each method. While all the considered methods imply a flat metallicity gradient, 
both the KK04 and KD02 calibrations provide metallicities that are larger than those from SSCs, while the PP04 calibration
give metallicities that are lower than ours. Figure~\ref{fig3} also shows that the M91-based metallicities better agree  
to the direct metallicity  determination. 
However, we stress that only the KK04 and M91 calibrations rely on the set of 
emission lines actually observed in \citet{bastian09} data, i.e. [\ion{O}{2}]$\lambda$3727, [\ion{O}{3}]$\lambda$4959,
[\ion{O}{3}]$\lambda$5007, H$\beta$; while both the PP04 and KD02 calibrations include diagnostic 
lines (i.e. [\ion{N}{2}]$\lambda$6583, H$\alpha$) not present in the original \citet{bastian09} spectra.
Moreover, we note that all the abundances from Figure~\ref{fig3} agree within $\pm$2$\sigma$.
As a result, we caution that the comparison 
presented in Figure~\ref{fig3} is purely qualitative, because we are not comparing actual measurements 
but using instead empirically derived metallicity conversions with their associated uncertainties.

The data suggest the presence of a flat abundance gradient, however
they sample only a small range in distances from the galaxy centre. 
Indeed, the number of SSCs observed should be increased considerably, 
especially in the inner regions, in order to test for radial variations and 
confirm whether the metallicity gradient is as shallow as found by \citet{bastian09} and suggested by our data.

\section{Summary}\label{SUMMARY}

Knowledge of the chemical composition of galaxies is fundamental
to trace back the history of cosmic chemical enrichment and understand 
the processes at work in galaxy formation and evolution.
Chemical abundances in extragalactic environments are commonly 
based on \ion{H}{2} region optical emission-line ratios. 
Numerous relations have been proposed to convert diagnostic emission line 
ratios into metallicity \citep[see for a review][]{kewley02}. Nonetheless, 
comparisons between abundances obtained from different calibrations show 
a systematic offset in metallicity estimates, that can amount up 
to $\sim$0.7 dex \citep{bresolin08,kewley08}.

We have developed a new method to avoid these calibration issues by performing 
quantitative spectroscopy of RSG stars or RSG-dominated SSCs in external galaxies. 
In this paper we use this $J$-band technique on KMOS spectra of three SSCs in NGC~4038
to measure precise metallicities \citep[see][]{davies10,gazakSSC,gazakPerOB1,gazak15,davies15,patrick15}
We find an average metallicity of [Z] = +0.07$\pm$ 0.03 ($\sigma$=0.05 dex).
Given the uncertainties, this is in good agreement with the results from \ion{H}{2} regions data by \citet{bastian09},
when the \citet{mc91} calibration is used to 
determine abundances. Furthermore, we find no evidence for a metallicity gradient.
However, a larger systematic study of RSGs is needed to assess 
the presence (or not) of a metallicity gradient among the young population within this merger galaxy pair. 

With the multi-object near IR spectrographs such as KMOS/VLT and MOSFIRE/Keck 
we can now investigate the chemical evolution of galaxies out to 
$\sim$7 Mpc from individual RSG stars (i.e. \citealp{davies10,davies15,patrick15,gazak15}).
Using the same technique on SSC, we here measure metallicity
as precise as ~0.10 dex out to the astonishing distance of $\sim$20Mpc in less than 
one night of observations. This opens new windows for extragalactic spectroscopy.
Indeed, the $J$-band method will allow us to quantitatively study the chemical evolution of 
galaxies --up to the Coma cluster -- in a way similar to current Galactic studies \citep{evans11}, when 
the next generation of extremely large telescopes, equipped with
adaptive optics supported near IR multi-object spectrographs, will be available to the community.

\acknowledgments
Based on observations made with ESO Telescopes at the La Silla Paranal Observatory under programme ID 093.B-0023.
We are very grateful to the anonymous referee who helped us improve our manuscript. 
RPK and JZG gratefully acknowledge support by the National Science foundation (NSF) under grant AST-1108906.
BP thanks the CNRS Programme National de Physique Stellaire for financial support.

{\it Facilities:} \facility{VLT(KMOS)}.

\clearpage


\begin{thebibliography}{}

 \bibitem[Allard et al.(2000)]{allard00} Allard, F., Hauschildt, P.~H., \& Schwenke, D.\ 2000, \apj, 540, 1005 
\bibitem[Andrews \& Martini(2013)]{andrews13} Andrews, B.~H., \& Martini, P.\ 2013, \apj, 765, 140 
\bibitem[Asplund et al.(2009)]{asplund09} Asplund, M., Grevesse, N., Sauval, A.~J., \& Scott, P.\ 2009, \araa, 47, 481 
\bibitem[Barnes \& Hernquist(1996)]{barnes06} Barnes, J.~E., \& Hernquist, L.\ 1996, \apj, 471, 115 
\bibitem[Barnes(1988)]{barnes88} Barnes, J.~E.\ 1988, \apj, 331, 699 
\bibitem[Baron \& White(1987)]{baron87} Baron, E., \& White, S.~D.~M.\ 1987, \apj, 322, 585 
\bibitem[Bastian et al.(2006)]{bastian06} Bastian, N., Emsellem, E., Kissler-Patig, M., \& Maraston, C.\ 2006, \aap, 445, 471 
\bibitem[Bastian et al.(2009)]{bastian09} Bastian, N., Trancho, G., Konstantopoulos, I.~S., \& Miller, B.~W.\ 2009, \apj, 701, 607 
\bibitem[Bergemann et al.(2012)]{bergemann12} Bergemann, M., Kudritzki, R.-P., Plez, B., et al.\ 2012, \apj, 751, 156 
\bibitem[Bergemann et al.(2013)]{bergemann13} Bergemann, M., Kudritzki, R.-P., W{\"u}rl, M., et al.\ 2013, \apj, 764, 115 
\bibitem[Bergemann et al.(2015)]{bergemann15} Bergemann, M., Kudritzki, R.-P., Gazak, Z., Davies, B., \& Plez, B.\ 2015, \apj, 804, 113 
\bibitem[Brandl et al.(2009)]{brandl09} Brandl, B.~R., Snijders, L., den Brok, M., et al.\ 2009, \apj, 699, 1982 
\bibitem[Bresolin et al.(2005)]{bresolin05} Bresolin, F., Schaerer, D., Gonz{\'a}lez Delgado, R.~M., \& Stasi{\'n}ska, G.\ 2005, \aap, 441, 981 
\bibitem[Bresolin(2008)]{bresolin08} Bresolin, F.\ 2008, The Metal-Rich Universe, 155 
\bibitem[Bresolin et al.(2009)]{bresolin09} Bresolin, F., Gieren, W., Kudritzki, R.-P., et al.\ 2009, \apj, 700, 309 
\bibitem[Bresolin et al.(2012)]{bresolin12} Bresolin, F., Kennicutt, R.~C., \& Ryan-Weber, E.\ 2012, \apj, 750, 122 
\bibitem[Bundy et al.(2009)]{bundy09} Bundy, K., Fukugita, M., Ellis, R.~S., et al.\ 2009, \apj, 697, 1369 
\bibitem[Chiappini et al.(2001)]{chiappini01} Chiappini, C.,Matteucci, F., \& Romano, D.\ 2001, \apj, 554, 1044 
\bibitem[Chien et al.(2007)]{chien07} Chien, L.-H., Barnes, J.~E., Kewley, L.~J., \& Chambers, K.~C.\ 2007, \apjl, 660, L105 
\bibitem[Cullen et al.(2007)]{cullen07} Cullen, H., Alexander, P., Green, D.~A., \& Sheth, K.\ 2007, \mnras, 376, 98 
\bibitem[Davies et al.(2009)]{davies09} Davies, B., Origlia, L., Kudritzki, R.-P., et al.\ 2009, \apj, 694, 46 
\bibitem[Davies et al.(2010)]{davies10} Davies, B., Kudritzki, R.-P., \& Figer, D.~F.\ 2010, \mnras, 407, 1203 
\bibitem[Davies et al.(2015)]{davies15} Davies, B., Kudritzki, R.-P., Gazak, Z., et al.\ 2015, \apj, 806, 21 
\bibitem[Davies et al.(2013)]{spark} Davies, R.~I., Agudo Berbel, A., Wiezorrek, E., et al.\ 2013, \aap, 558, AA56 
\bibitem[de Ravel et al.(2009)]{deravel09} de Ravel, L., Le F{\`e}vre, O., Tresse, L., et al.\ 2009, \aap, 498, 379 
\bibitem[Edmunds \& Pagel(1984)]{edmunds84} Edmunds, M.~G., \& Pagel, B.~E.~J.\ 1984, \mnras, 211, 507 
\bibitem[Ellison et al.(2008)]{ellison08} Ellison, S.~L., Patton, D.~R., Simard, L., \& McConnachie, A.~W.\ 2008, \aj, 135, 1877 
\bibitem[Emonts et al.(2006)]{emonts06} Emonts, B.~H.~C., Morganti, R., Tadhunter, C.~N., et al.\ 2006, \aap, 454, 125 
\bibitem[Ercolano et al.(2010)]{ercolano10} Ercolano, B., Wesson, R., \& Bastian, N.\ 2010, \mnras, 401, 1375 
\bibitem[Evans et al.(2011)]{evans11} Evans, C.~J., Davies, B., Kudritzki, R.-P., et al.\ 2011, \aap, 527, AA50 
\bibitem[Fabbiano et al.(2004)]{fabbiano04} Fabbiano, G., Baldi, A., King, A.~R., et al.\ 2004, \apjl, 605, L21 
\bibitem[Fu et al.(2009)]{fu09} Fu, J., Hou, J.~L., Yin, J., \& Chang, R.~X.\ 2009, \apj, 696, 668 
\bibitem[Garnett et al.(1997)]{garnett97} Garnett, D.~R.,Shields, G.~A., Skillman, E.~D., Sagan, S.~P., \& Dufour, R.~J.\ 1997, \apj, 489, 63 
\bibitem[Gazak et al.(2013)]{gazakM83} Gazak, J.~Z., Bastian, N., Kudritzki, R.-P., et al.\ 2013, \mnras, 430, L35 
\bibitem[Gazak et al.(2014a)]{gazakPerOB1} Gazak, J.~Z., Davies, B., Kudritzki, R., Bergemann, M., \& Plez, B.\ 2014a, \apj, 788, 58 
\bibitem[Gazak et al.(2014b)]{gazakSSC} Gazak, J.~Z., Davies, B., Bastian, N., et al.\ 2014b, \apj, 787, 142 
\bibitem[Gazak et al.(2015)]{gazak15} Gazak, J.~Z., Kudritzki, R., Evans, C., et al.\ 2015, \apj, 805, 182 
\bibitem[Georgakakis et al.(2000)]{georg00} Georgakakis, A., Forbes, D.~A., \& Norris, R.~P.\ 2000, \mnras, 318, 124 
\bibitem[Gilbert \& Graham(2007)]{gilbert07} Gilbert, A.~M., \& Graham, J.~R.\ 2007, \apj, 668, 168 
\bibitem[Gustafsson et al.(2008)]{gustafsson08} Gustafsson, B., Edvardsson, B., Eriksson, K., et al.\ 2008, \aap, 486, 951 
\bibitem[Hibbard \& van Gorkom(1996)]{hibbard96} Hibbard, J.~E., \& van Gorkom, J.~H.\ 1996, \aj, 111, 655 
\bibitem[Hibbard et al.(2005)]{hibbard05} Hibbard, J.~E., Bianchi, L., Thilker, D.~A., et al.\ 2005, \apjl, 619, L87 
\bibitem[Hosek et al.(2014)]{hosek14} Hosek, M.~W., Jr., Kudritzki, R.-P., Bresolin, F., et al.\ 2014, \apj, 785, 151 
\bibitem[Humphreys \& Davidson(1979)]{humphreys79} Humphreys, R.~M., \& Davidson, K.\ 1979, \apj, 232, 409 
\bibitem[Iono et al.(2005)]{iono05} Iono, D., Yun, M.~S., \& Ho, P.~T.~P.\ 2005, \apjs, 158, 1 
\bibitem[Karachentsev et al.(2000)]{kara00} Karachentsev, I.~D., Sharina, M.~E., \& Huchtmeier, W.~K.\ 2000, \aap, 362, 544 
\bibitem[Karl et al.(2010)]{karl10} Karl, S.~J., Naab, T., Johansson, P.~H., et al.\ 2010, \apjl, 715, L88 
\bibitem[Kennicutt et al.(1987)]{kennicutt87} Kennicutt, R.~C., Jr., Roettiger, K.~A., Keel, W.~C., van der Hulst, J.~M., \& Hummel, E.\ 1987, \aj, 93, 1011 
\bibitem[Kewley \& Dopita(2002)]{kewley02} Kewley, L.~J., \& Dopita, M.~A.\ 2002, \apjs, 142, 35 
\bibitem[Kewley \& Ellison(2008)]{kewley08} Kewley, L.~J., \& Ellison, S.~L.\ 2008, \apj, 681, 1183 
\bibitem[Kewley et al.(2006)]{kewley06} Kewley, L.~J., Geller, M.~J., \& Barton, E.~J.\ 2006, \aj, 131, 2004 
\bibitem[Kewley et al.(2010)]{kewley10} Kewley, L.~J., Rupke, D., Zahid, H.~J., Geller, M.~J., \& Barton, E.~J.\ 2010, \apjl, 721, L48 
\bibitem[Klaas et al.(2010)]{klaas10} Klaas, U., Nielbock, M., Haas, M., Krause, O., \& Schreiber, J.\ 2010, \aap, 518, LL44 
\bibitem[Krabbe et al.(2008)]{krabbe08} Krabbe, A.~C., Pastoriza, M.~G., Winge, C., Rodrigues, I., \& Ferreiro, D.~L.\ 2008, \mnras, 389, 1593 
\bibitem[Krabbe et al.(2011)]{krabbe11} Krabbe, A.~C., Pastoriza, M.~G., Winge, C., et al.\ 2011, \mnras, 416, 38 
\bibitem[Kudritzki et al.(2012)]{kudritzki12} Kudritzki, R.-P., Urbaneja, M.~A., Gazak, Z., et al.\ 2012, \apj, 747, 15 
\bibitem[Kudritzki et al.(2013)]{kudritzki13} Kudritzki, R.-P., Urbaneja, M.~A., Gazak, Z., et al.\ 2013, \apjl, 779, LL20 
\bibitem[Kudritzki et al.(2014)]{kudritzki14} Kudritzki, R.-P., Urbaneja, M.~A., Bresolin, F., Hosek, M.~W., Jr.,\& Przybilla, N.\ 2014, \apj, 788, 56 
\bibitem[Kudritzki et al.(2015)]{kudritzki15} Kudritzki, R.-P., Ho, I.-T., Schruba, A., et al.\ 2015, \mnras, 450, 342 
\bibitem[Kobulnicky \& Kewley(2004)]{kobu04} Kobulnicky, H.~A., \& Kewley, L.~J.\ 2004, \apj, 617, 240 
\bibitem[Lauberts \& Valentijn(1989)]{lauberts89} Lauberts, A., \& Valentijn, E.~A.\ 1989, The Messenger, 56, 31 
\bibitem[Lee et al.(2004)]{lee04} Lee, J.~C., Salzer, J.~J., \& Melbourne, J.\ 2004, \apj, 616, 752 
\bibitem[Lequeux et al.(1979)]{lequeux79} Lequeux, J., Peimbert, M., Rayo, J.~F., Serrano, A., \& Torres-Peimbert, S.\ 1979, \aap, 80, 155 
\bibitem[Maiolino et al.(2008)]{maiolino08} Maiolino, R., Nagao, T., Grazian, A., et al.\ 2008, \aap, 488, 463 
\bibitem[Martin(2006)]{martin06} Martin, C.~L.\ 2006, \apj, 647, 222 
\bibitem[McGaugh(1991)]{mc91} McGaugh, S.~S.\ 1991, \apj, 380, 140 
\bibitem[McGaugh(1994)]{mcgaugh94} McGaugh, S.~S.\ 1994, \apj, 426, 135 
\bibitem[Mengel et al.(2002)]{mengel02} Mengel, S., Lehnert, M.~D., Thatte, N., \& Genzel, R.\ 2002, \aap, 383, 137 
\bibitem[Meyer(1985)]{meyer85} Meyer, J.-P.\ 1985, \apjs, 57, 173 
\bibitem[Michel-Dansac et al.(2008)]{michel08} Michel-Dansac, L., Lambas, D.~G., Alonso, M.~S., \& Tissera, P.\ 2008, \mnras, 386, L82 
\bibitem[Mihos \& Hernquist(1996)]{mihos96} Mihos, J.~C., \& Hernquist, L.\ 1996, \apj, 464, 641 
\bibitem[Mihos et al.(1993)]{mihos93} Mihos, J.~C., Bothun, G.~D., \& Richstone, D.~O.\ 1993, \apj, 418, 82 
\bibitem[Mirabel et al.(1998)]{mirabel98} Mirabel, I.~F., Vigroux, L., Charmandaris, V., et al.\ 1998, \aap, 333, L1 
\bibitem[Mott et al.(2013)]{mott13} Mott, A., Spitoni, E., \& Matteucci, F.\ 2013, \mnras, 435, 2918 
\bibitem[Pagel et al.(1979)]{pagel79} Pagel, B.~E.~J., Edmunds, M.~G., Blackwell, D.~E., Chun, M.~S., \& Smith, G.\ 1979, \mnras, 189, 95 
\bibitem[Patrick et al.(2015)]{patrick15} Patrick, L.~R., Evans, C.~J., Davies, B., et al.\ 2015, \apj, 803, 14 
\bibitem[Pettini \& Pagel(2004)]{pettini04} Pettini, M., \& Pagel, B.~E.~J.\ 2004, \mnras, 348, L59 
\bibitem[Pilkington et al.(2012)]{pilkington12} Pilkington, K., Few, C.~G., Gibson, B.~K., et al.\ 2012, \aap, 540, A56 
\bibitem[Portegies Zwart et al.(2010)]{portgeis10}Portegies Zwart, S.~F., McMillan, S.~L.~W., \& Gieles, M.\ 2010, \araa, 48, 431 
\bibitem[Prantzos \& Boissier(2000)]{prantzos00} Prantzos, N., \& Boissier, S.\ 2000, \mnras, 313, 338 
\bibitem[Rampazzo et al.(2005)]{rampazzo05} Rampazzo, R., Plana, H., Amram, P., et al.\ 2005, \mnras, 356, 1177 
\bibitem[Rich et al.(2012)]{rich12} Rich, J.~A., Torrey, P., Kewley, L.~J., Dopita, M.~A., \& Rupke, D.~S.~N.\ 2012, \apj, 753, 5 
\bibitem[Riess et al.(2011)]{riess11} Riess, A.~G., Macri, L., Casertano, S., et al.\ 2011, \apj, 730, 119 
\bibitem[Rosa et al.(2014)]{rosa14} Rosa, D.~A., Dors, O.~L., Krabbe, A.~C., et al.\ 2014, \mnras, 444, 2005 
\bibitem[Rubin et al.(1994)]{rubin94} Rubin, R.~H., Simpson, J.~P., Lord, S.~D., et al.\ 1994, \apj, 420, 772 
\bibitem[Rupke et al.(2005)]{rupke05} Rupke, D.~S., Veilleux, S., \& Sanders, D.~B.\ 2005, \apjs, 160, 115 
\bibitem[Rupke et al.(2008)]{rupke08} Rupke, D.~S.~N., Veilleux, S., \& Baker, A.~J.\ 2008, \apj, 674, 172 
\bibitem[S{\'a}nchez et al.(2014)]{sanchez14} S{\'a}nchez, S.~F., Rosales-Ortega, F.~F., Iglesias-P{\'a}ramo, J., et al.\ 2014, \aap, 563, AA49 
\bibitem[Skillman(1989)]{skillman89} Skillman, E.~D.\ 1989, \apj, 347, 883 
\bibitem[Searle(1971)]{searle71} Searle, L.\ 1971, \apj, 168, 327 
\bibitem[Sharples et al.(2013)]{sharples13} Sharples, R., Bender, R., Agudo Berbel, A., et al.\ 2013, The Messenger, 151, 21 
\bibitem[Stasi{\'n}ska(2005)]{stasinska05} Stasi{\'n}ska, G.\ 2005, \aap, 434, 507 
\bibitem[Teyssier et al.(2010)]{teyssier10} Teyssier, R., Chapon, D., \& Bournaud, F.\ 2010, \apjl, 720, L149 
\bibitem[Thim et al.(2003)]{thim03} Thim, F., Tammann, G.~A., Saha, A., et al.\ 2003, \apj, 590, 256 
\bibitem[Toomre \& Toomre(1972)]{toomre72} Toomre, A., \& Toomre, J.\ 1972, \apj, 178, 623 
\bibitem[Trancho et al.(2007)]{trancho07} Trancho, G., Bastian, N., Miller, B.~W., \& Schweizer, F.\ 2007, \apj, 664, 284 
\bibitem[Tremonti et al.(2004)]{tremonti04} Tremonti, C.~A., Heckman, T.~M., Kauffmann, G., et al.\ 2004, \apj, 613, 898 
\bibitem[Torres-Flores et al.(2014)]{torres14} Torres-Flores, S., Scarano, S., Mendes de Oliveira, C., et al.\ 2014, \mnras, 438, 1894 
\bibitem[Vacca \& Conti(1992)]{vacca92} Vacca, W.~D., \& Conti, P.~S.\ 1992, \apj, 401, 543 
\bibitem[Werk et al.(2011)]{werk11} Werk, J.~K., Putman, M.~E., Meurer, G.~R., \& Santiago-Figueroa, N.\ 2011, \apj, 735, 71 
\bibitem[Wilson et al.(2000)]{wilson00} Wilson, C.~D., Scoville, N., Madden, S.~C., \& Charmandaris, V.\ 2000, \apj, 542, 120 
\bibitem[Whitmore \& Schweizer(1995)]{whitmore95} Whitmore, B.~C., \& Schweizer, F.\ 1995, \aj, 109, 960 
\bibitem[Whitmore et al.(1999)]{whitmore99} Whitmore, B.~C., Zhang, Q., Leitherer, C., et al.\ 1999, \aj, 118, 1551 
\bibitem[Whitmore et al.(2010)]{whitmore10} Whitmore, B.~C., Chandar, R., Schweizer, F., et al.\ 2010, \aj, 140, 75 
\bibitem[Whitmore et al.(2014)]{whitmore14} Whitmore, B.~C., Brogan, C., Chandar, R., et al.\ 2014, \apj, 795, 156 
\bibitem[Zaritsky et al.(1994)]{zaritsky94} Zaritsky, D., Kennicutt, R.~C., Jr., \& Huchra, J.~P.\ 1994, \apj, 420, 87 
\bibitem[Zhang et al.(2001)]{zhang01} Zhang, Q., Fall, S.~M., \& Whitmore, B.~C.\ 2001, \apj, 561, 727 
\bibitem[Zurita \& Bresolin(2012)]{zurita12} Zurita, A., \& Bresolin, F.\ 2012, \mnras, 427, 1463 



\end{thebibliography}
\end{document}